%% file: sptpol_des.tex
\documentclass[twocolumn]{aastex61}

\newcommand{\ApJsubmit}{}
\newcommand{\allauthorsinfront}{}

\usepackage{multirow}
\usepackage{color}
\usepackage{amsmath,amsfonts,bm}
\usepackage{hyperref}
\usepackage[T1]{fontenc}
\usepackage{graphicx}	
\usepackage{dcolumn}
\usepackage{lineno}
\usepackage{seqsplit}

\newcolumntype{C}[1]{>{\centering\let\newline\\\arraybackslash\hspace{0pt}}m{#1}}

\newcommand{\refresponse}[1]{\textcolor{black}{#1}}

\newcommand{\sptpol}{SPTpol} 
\newcommand{\sptsz}{SPT-SZ} 
\newcommand{\planck}{{\it Planck}}
\newcommand{\snr}{$S/N$}

\newcommand{\fitAforvlsamplewitherrors}{$2.70^{+0.51}_{-0.50}$}
\newcommand{\fitAforvlsamplewitherrorsformatted}{\ensuremath{2.70 \pm 0.51}}

\newcommand{\howmanyrandomsforMF}{50,000}
\newcommand{\howmanysigmaforfullsample}{8.7$\sigma$}
\newcommand{\howmanysigmaforcosmosample}{6.7$\sigma$}
\newcommand{\howmanyclustersinfullsample}{4003}
\newcommand{\howmanyclustersincosmosample}{1741}

\newcommand{\fitmassforfullsamplewitherrors}{$1.28^{+0.14}_{-0.18}$}
\newcommand{\fitmassforfullsamplewithsyserrors}{\ensuremath{1.28^{+0.14}_{-0.18} {\rm ~[stat.]} \pm 0.03 ~{\rm [sys.]} }} 

\newcommand{\fitmassforvlsamplewitherrors}{$1.62^{+0.32}_{-0.25}$}
\newcommand{\fitmassforvlsamplewithsyserrors}{\ensuremath{1.62^{+0.32}_{-0.25}{\rm ~[stat.]} \pm 0.04 ~{\rm [sys.]}}}

\newcommand{\mvir}{\ensuremath{M_{\rm 200m}}}
\newcommand{\kappaonehalofull}{$\kappa^{1h}(\theta)$}

\newcommand{\kappaonehalomz}{${\kappa}^{1h}(M,z)$}
\newcommand{\kappatwohalomz}{${\kappa}^{2h}(M, z)$}
\newcommand{\kappatotalmz}{${\kappa}(M, z)$}

\newcommand{\kappaonehalo}{${\kappa}^{1h}$}
\newcommand{\kappatwohalo}{${\kappa}^{2h}$}

\newcommand{\boxsize}{$300^{\prime} \times 300^{\prime}$}

\newcommand{\whichyear}{\mbox{Year-3}}
\newcommand{\whichcatversion}{\texttt{y3\_gold:v6.4.22}}

\newcommand{\ML}{\ensuremath{M-\lambda}}

\newcommand{\am}{$^{\prime}$}
\newcommand{\ukam}{${\rm \mu K^{\prime}}$ }
\newcommand{\sqdeg}{deg$^{2}$}
\newcommand{\tszfreemapnotation}{{\rm SZ{\text -}free}}

\newcommand{\bnhat}{\bm{\hat{\mbox{n}}}}
\newcommand{\bL}{\bm{\ell}}

\newcommand{\RM}{redMaPPer} 
\newcommand{\msol}{\ensuremath{{M}_{\odot}}}
\newcommand{\munits}{\ensuremath{10^{14}~ \msol}}

\newcommand{\lcdm}{$\Lambda$CDM}
\newcommand{\beambl}{$B_{\ell}$}
\newcommand{\comment}[1]{}

\defcitealias{raghunathan17a}{R17}
\defcitealias{baxter18}{B18}
\defcitealias{sehgal10}{S10}
\defcitealias{mcclintock18}{M18}
\begin{document}

\begin{nolinenumbers}
\vspace*{-\headsep}\vspace*{\headheight}
\footnotesize \hfill FERMILAB-PUB-17-524-AE\\
\vspace*{-\headsep}\vspace*{\headheight}
\footnotesize \hfill DES-2018-0388
\end{nolinenumbers}

\ifdefined\allauthorsinfront
\else
\AuthorCallLimit=12
\fi

\title{Mass Calibration of Optically Selected DES clusters using a Measurement of CMB-Cluster Lensing with \sptpol{} Data}

\shorttitle{CMB-cluster lensing with {\rm \sptpol{}} and DES datasets}
\shortauthors{S. Raghunathan, S. Patil \it{et al.}}

\correspondingauthor{Srinivasan Raghunathan}
\email{sri@physics.ucla.edu}


\ifdefined\ApJsubmit
\input{principal_authors}

\input{spt_des_authorlist}

\else

\input{author_lists/principal_authors}
\input{author_lists/spt_des_authorlist}
\fi



\date{Accepted 22 January 2019. Received 2 January 2019; in original form 26 October 2018}

\begin{abstract}
We use cosmic microwave background (CMB) temperature maps from the 500\,\sqdeg{} \sptpol{} survey to measure the stacked lensing convergence of galaxy clusters from the Dark Energy Survey (DES) \whichyear{} \RM{} (RM) cluster catalog. 
The lensing signal is extracted through a modified quadratic estimator designed to be unbiased by the thermal Sunyaev-Zel{'}dovich (tSZ) effect. 
The modified estimator uses a tSZ-free map, constructed from the \sptpol{} 95 and 150 GHz datasets, to estimate the background CMB gradient. 
For lensing reconstruction, we employ two versions of the RM catalog: a flux-limited sample containing 4003 clusters and a volume-limited sample with 1741 clusters. 
We detect lensing at a significance of \howmanysigmaforfullsample(\howmanysigmaforcosmosample) with the flux(volume)-limited sample. 
By modeling the reconstructed convergence using the Navarro-Frenk-White profile, we find the average lensing masses to be \mbox{$\mvir = (\fitmassforvlsamplewithsyserrors)$} and \mbox{$(\fitmassforfullsamplewithsyserrors)$ $\times \munits$} for the volume- and flux-limited samples respectively. 
The systematic error budget is much smaller than the statistical uncertainty and is dominated by the uncertainties in the RM cluster centroids. 
We use the volume-limited sample to calibrate the normalization of the mass-richness scaling relation, and find a result consistent with the galaxy weak-lensing measurements from DES \citep{mcclintock18}.
\end{abstract}



\section{Introduction}\label{sec_intro}

The abundance of galaxy clusters as a function of mass and redshift is highly sensitive to the the details of structure growth and the geometry of the Universe \citep[e.g.,][]{allen11}. 
Past cluster surveys have yielded competitive constraints on a number of open questions in cosmology today, most notably on the sum of the neutrino masses and the drivers for cosmic acceleration \citep{mantz08, vikhlinin09, rozo10, hasselfield13, mantz15, placksz15, dehaan16, salvati17}. 
Future surveys \citep{lsst09, erosita12, benson14, henderson16, cmbs4-sb1, SO18} will find tens to hundreds of thousands of galaxy clusters, with the potential for significantly better cosmological constraints. 
Achieving this improvement, however, will also require a calibration between cluster mass with observable quantities such as X-ray luminosity, the Sunyaev-Zel{'}dovich (SZ) effect, or optical richness \citep{sunyaev72,sunyaev80b, vikhlinin09b, rozo10, appelgate14, linden14}. 

Gravitational lensing is one of the most promising techniques to estimate galaxy cluster masses.
Gravitational lensing has the significant advantage that it directly probes the total matter distribution in a galaxy cluster, without depending on complex baryonic physics. 
Optical weak-lensing measurements have demonstrated accurate mass estimates which have been used in recent cluster cosmological analyses \citep{rozo13, linden14}.
While galaxies may be the most well-known lensing source \citep{henk13}, any background light source can be used. 
The cosmic microwave background (CMB) is an effective alternative due to its extremely well measured statistical properties and known high redshift ($z\sim1100$). 
CMB-cluster lensing is particularly powerful for high-redshift clusters for which it is more difficult to observe background galaxies with sufficient signal-to-noise (\snr). 
Consequently, it is one of the most promising methods for future CMB surveys including CMB-S4 that are expected to return thousands of high redshift ($z>1$) clusters \citep{cmbs4-sb1}. 
For low-redshift clusters, CMB lensing is complementary to galaxy weak-lensing measurements, as the systematics associated with the two measurements are different.
However, the CMB-cluster lensing signal is small. 
We estimate the lensing \snr{} to be $\sim0.5$ for a cluster with \mbox{M $\sim$ \munits{}} even for a futuristic experiment 
like CMB-S4.
So we are limited to measuring the average mass of a set of clusters.

Several estimators have been proposed to extract the CMB-cluster lensing signal using CMB temperature and polarization maps \citep{seljak00b, dodelson04, holder04, maturi05, lewis06, hu07, yoo08, yoo10, melin15, horowitz17}.
Measurements have now been performed by a number of experiments using CMB temperature data. 
\citet{baxter15} detected CMB-cluster lensing at 3.1-sigma using South Pole Telescope (SPT) SZ survey data for a sample 513 SPT-selected clusters.
Additional detections of CMB-cluster lensing have been made using ACTPol \citep{matthew15} and \planck{} data \citep{placksz15, raghunathan17b}.
CMB-cluster lensing has also been used to calibrate the mass-richness (\ML) relation of the \RM{} (RM) algorithm using both \planck{} data at the locations of clusters in Sloan Digital Sky Survey \citep[SDSS,][]{geach17} data, and SPT-SZ data at the locations of clusters in Dark Energy Survey (DES) Year-1 \citep[hereafter \citetalias{baxter18}]{baxter18} data.

These initial measurements have estimated the lensing signal from CMB temperature data.
Lensing measurements using temperature data are susceptible to bias from foreground signals, in particular the thermal SZ (tSZ) signal from the cluster itself. 
The bias due to the tSZ effect can be mitigated by using tSZ-free maps for lensing measurements \citep{baxter15} or by including additional filtering when estimating the background gradient with a lensing quadratic estimator \citepalias[QE,][]{baxter18}. 
Both of these methods reduce the lensing \snr. 
We follow a different strategy here by reworking the standard QE to use a tSZ-free gradient map from the \sptpol{} survey for a tSZ-bias free lensing reconstruction.
While this paper was in the production stage, \citet{madhavacheril18} published a similar method using simulated datasets where they also demonstrated that the tSZ-free gradient quadratic estimators can robustly reconstruct CMB lensing using temperature data alone.

In the current work, we apply the modified QE to \sptpol{} CMB temperature maps and reconstruct the lensing signal at the location of galaxy clusters from the DES \whichyear{} RM catalog. 
We employ two samples of the RM catalog and obtain lensing detection significances of \howmanysigmaforfullsample{} with \howmanyclustersinfullsample{}  clusters from the flux-limited sample and \howmanysigmaforcosmosample{} for a smaller volume-limited sample containing \howmanyclustersincosmosample{} clusters.
We use the lensing measurements from the volume-limited sample to calibrate the \ML{} relation of the RM cluster sample at the 18\% level.
We validate our results against several sources of systematic errors and note that the uncertainty in the knowledge of the cluster mis-centering introduces $\sim3\%$ error in our lensing measurements, which is sub-dominant compared to the statistical error.

The paper is organized as follows: In \S\ref{sec_data} we describe the \sptpol{} CMB temperature map and the DES RM cluster catalog. 
This is followed by a description of the lensing estimator, simulations used to validate the pipeline, cluster convergence profiles, cutout extraction, and the modeling in \S\ref{sec_methods}. 
Pipeline and data validation along with the estimates of the systematic error budgets are summarized in \S\ref{sec_validation}.
We present our lensing measurements and compare them to literature in \S\ref{sec_results}.
The conclusion is in \S\ref{sec_conclusion}. 

Throughout this work, we use the best-fit $\Lambda$CDM cosmology obtained from the chain that combines {\it Planck} 2015 data with external datasets \path{TT,TE,EE+lowP+lensing+ext} \citep{planck15-13}. 
We define all halo quantities with respect to the radius $R_{\rm 200m}$ defined as the region within which the average mass density is 200 times the mean density of the universe at the halo redshift.
For parameter constraints, we report the median values and $1\sigma$ uncertainties from the 16$^{th}$ and 84$^{th}$ percentiles.


\section{Data}\label{sec_data}
We describe the CMB datasets from the \sptpol{} survey in \S\ref{sec_sptpol}. 
This is followed by a brief description of the DES experiment and the selection of the cluster catalog used in this work in \S\ref{sec_DES}.

\subsection{\sptpol{} {\rm 500} deg$^{2}$ survey}\label{sec_sptpol}

\sptpol{} is the second camera installed on the \mbox{10-meter} South Pole Telescope \citep[SPT,][]{padin08, carlstrom11} located at the Amundsen-Scott South Pole station.
The \sptpol{} focal plane consists of 1536 polarization-sensitive transition edge sensor bolometers (360 at 95 GHz and 1176 at 150 GHz) \citep{austermann12}.
The \sptpol{} 500~deg$^{2}$ survey spans fifteen degrees of declination, from -65 to -50 degrees, and four hours of right ascension, from 22h to 2h. 
In this work, we use CMB temperature maps from observations between April 2013 and September 2016 in frequency bands centered at approximately 95 GHz and 150 GHz. 
The telescope beam and pointing solutions were characterized using Venus and bright point sources in the \sptpol{} survey region. 
The final telescope beam along with the pointing jitter roughly corresponds to a $\theta_{\rm FWHM} = 1.^{\prime}22\ (1.^{\prime}7)$ Gaussian for the 150 (95) GHz dataset. 

We briefly summarize the procedure we use to reduce raw CMB data to maps and refer the reader to \cite{henning18} for further details. 
The raw data are composed of digitized time-ordered data (TOD) for each detector that are converted into CMB temperature units. 
We bin the TOD into two different maps using a flat-sky approximation in the Sanson-Flamsteed projection \citep{calabretta02, schaffer11}. 
To construct the first map, in which we aim to reconstruct the small-scale lensing signal, we remove large-scale modes $\ell \le 300$, bandpass filter the TOD in the range of approximately $300 \le \ell_{x} \le 20,000$, and bin them into 0.\am5\ square pixels.
For the second map, intended for estimation of the large-scale CMB gradient, we apply minimal TOD filtering by only removing modes below $\ell_{x} \le 30$, and bin them into 3\am\ square pixels.
While we only use the data from the 150 GHz channel for the first map, the latter is a tSZ signal cleaned map produced by linearly combining the 95 and 150 GHz channels. 
We use this tSZ-free map to reconstruct the background gradient of the CMB at the cluster locations.
As we will see later in \S\ref{sec_method_QE}, the gradient estimation using the tSZ-free map helps in removing the tSZ-induced lensing bias.
The minimal filtering on this map allows us to recover large-scale modes which indeed helps in a better estimation of the background gradient.
The 0.\am5{} resolution 150 GHz map has a white noise level of \mbox{$\Delta_{\rm T} = $ 6 \ukam} estimated using a jackknife approach. 
The low-resolution tSZ-free combination is noisier with \mbox{$\Delta_{\rm T} \sim $ 17 \ukam}.

\subsection{DES and the {\rm \RM} catalog}\label{sec_DES}
The Dark Energy Survey (DES) is a $\sim$5000 \sqdeg, optical to near-infrared survey conducted using the Dark Energy Camera \citep{flaugher15} mounted on the 4-meter Victor Blanco telescope at Cerro Tololo Observatory in Chile and has recently begun its sixth year of observations. 
For this analysis, we use the cluster catalog obtained from the first three years of DES observations, which almost covers the \sptpol{} 500\,deg$^{2}$ survey. 

The cluster catalog was derived using the RM algorithm \citep{rykoff14}.
RM is an optical cluster-finding algorithm which detects candidates by identifying over-densities of luminous red galaxies with luminosity greater than 20\% of $L_{*}.$
It is based on our understanding that galaxy clusters are agglomerations of galaxies containing old and subsequently red stars. 
The algorithm iteratively assigns membership and centering probabilities for each red galaxy identified as belonging to a cluster candidate. 
A weighted sum of the membership probabilities, richness $\lambda$, is assigned to each candidate.
The centre comes from the galaxy with the highest centering probability.
The DES RM catalog contains two samples: a flux-limited sample and a volume-limited sample. 
The flux-limited sample has more high-redshift clusters detected from deep fields in the survey. 
On the other hand, the volume-limited sample is independent of survey depth, complete above a luminosity threshold \citep[hereafter \citetalias{mcclintock18}]{mcclintock18}, and normally preferred for cosmological analysis.
See \citet{rykoff16} for more information on the application of RM to the DES survey data.

The RM cluster catalog version employed in this analysis is \whichcatversion. 
The \whichyear{} gold catalogue is based on the previous catalog from the Year-1 data \citep{drlica-wagner17} with some updates described in \citet{morganson18}.
The catalog contains 54,112 clusters above richness $\lambda \ge 20$ in the flux-limited sample and 21,094 clusters in the volume-limited sample. 
Of these, 5,828 (2,428) clusters from the flux(volume)-limited sample lie within the \sptpol{} 500 \sqdeg{} survey  in the redshift range $0.1 \le z \le$ 0.95 (0.90). 
We additionally remove clusters near the survey edges by removing the cutouts (see \ref{sec_cluster_cutouts}) with more than 5\% masked pixels or within $10^{\prime}$ distance from any bright ($\ge 6$ mJy at 150\,GHz) point sources detected in the \sptpol{} temperature map.
These cuts leave 4,003 (1,741) clusters with $\lambda \ge 20$ from the flux(volume)-limited sample with a median redshift of $\tilde{z}$ = 0.77 (0.48). 
The error in the cluster photo-$z$ estimates are small with $\hat{\sigma}_{z} = 0.01 (1+z)$ \citep{rozo15}.


\section{Methods}
\label{sec_methods}
We now turn to the method for measuring the cluster lensing signal. 
First, we describe the modified QE, which uses a tSZ-free gradient map to eliminate the tSZ-induced bias in \S\ref{sec_method_QE}.
Next, we present the lensing pipeline starting with the simulations used in the analysis in \S\ref{sec_sims_funda}, calculation of the cluster convergence profiles in \S\ref{sec_clus_profile}, cluster cutouts extraction in \S\ref{sec_cluster_cutouts}, the weighting scheme applied to obtain the stacked convergence in  \S\ref{subsec_weights}, and modeling in \S\ref{sec_nfw_model_fitting}.

\subsection{Quadratic estimator}
\label{sec_method_QE}

We use a quadratic lensing estimator \citep{hu07} to extract the cluster lensing signal.
Specifically, we obtain the convergence $\kappa$ which is related to the underlying lensing potential $\phi$ as $2\kappa = -\nabla^{2}\phi$.
The QE uses two maps to reconstruct the lensing convergence: one map of the CMB gradient on large scales, and one map of the CMB temperature fluctuations on small scales.
In the absence of lensing, the two maps would be uncorrelated. 
The convergence reconstructed from the two maps will be \citep{hu07}:
\begin{equation}
\hat{\kappa}_{\bL} = -A_{\ell} \int d^{2}\bnhat\ e^{-i\bnhat\cdot\bL}\ {\rm Re} \left\{ \nabla \cdot \left[ G(\bnhat) L^{*}(\bnhat)\right]\right\},
\label{eq_QE_kappa}
\end{equation} 
where $G$ is the temperature gradient map and $L$ is the temperature fluctuation map, both optimally filtered to maximize the lensing \snr. 
The two maps and the optimal weights are described in the next section.
The normalization factor $A_{\ell}$ can be calculated following Eq. (18) of \cite{hu07}.
Since the desired input to the QE is the gradient of the unlensed CMB, the gradient map $G$ is low-pass filtered (LPF) at $\ell_{\rm G}$ 
\citep{hu07} to avoid multipoles where the cluster lensing or foregrounds begin to enter. 
The LPF negligibly degrades the lensing \snr{} since most of the gradient information is at large scales (see Fig. 1 of \citealt{hu07}).

When, as in this work, temperature maps are used in both legs of the QE, lensing is not the only process that introduces correlations between the maps $G$ and $L$. 
Undesired correlations are also sourced by clusters' own SZ signals; these correlations lead to severe contamination of the lensing reconstruction. 
An obvious way to reduce the tSZ bias would be to generate a tSZ-free map from a linear combination of single-frequency maps; this has been done in previous analyses \citep{baxter15}. 
However this linear combination can substantially increase the map noise and degrade the lensing \snr{}. 
For instance, the tSZ-free map used by \citet{baxter15} had a noise level approximately three times higher than the SPT-SZ 150GHz map alone.
Modeling the tSZ signal is possible in principle as an alternative,  but we do not yet have an adequate understanding of the intracluster medium to do so reliably. 

Modifying the LPF in the gradient map,  $\ell_{\rm G}$, is another plausible alternative to reduce but not eliminate this correlation.
The lensing bias due to this correlation will be particularly large for massive nearby clusters that span a large angular extent on the sky.
While reducing the bias, adopting a lower $\ell_{\rm G}$ will reduce the number of modes for the gradient estimation and result in a lower \snr{}.
Thus, the choice of $\ell_{\rm G}$ is a trade-off between \snr{} and biases due to both the magnification effect considered by \citet{hu07} and from  foreground emission. 
For example, 
\citetalias{baxter18}, using the \sptsz{} temperature maps ($\Delta_{\rm T} = 18$\ukam{}), 
chose $\ell_{\rm G} = 1500$ and
reported an upper limit of $11\%$ on the tSZ-induced bias due to clusters in the richness range $\lambda \in [20,40]$.

\subsubsection{tSZ-free map for gradient estimation}
\label{subsec_QE_modificaltion}
A key point in this analysis is that for 
QE-based lensing reconstruction, we only need to eliminate tSZ-induced correlations between the  maps $G$ and $L$ used in the two legs, which can be done by removing the tSZ signal from either one of the maps.
Hence, instead of treating $\ell_{\rm G}$ as a free parameter used to reduce the tSZ-bias, we eliminate the bias completely by working with a tSZ-free map, $T^{\tszfreemapnotation}$, for the gradient estimation $G$.
Recently, \citet{madhavacheril18} also made a successful demonstration of this method independently using simulations.
In this analysis, the $T^{\tszfreemapnotation}$ map $G$ is a linear combination of the \sptpol{} 95 and 150\,GHz temperature data. 
The second map $L$ is the lower-noise \sptpol{} 150\,GHz data, $T^{150}$, alone. 

We can now write down expressions for the two maps, $G$ and $L$:
\begin{eqnarray}
G_{\bL} &=& i\bL{} \,W_{\ell}^{G}\, T^{\tszfreemapnotation}_{\bL},\\
L_{\bL} &=& W_{\ell}^{L} \,T^{150}_{\bL}.
\label{eq_QE_filtered_gradient_lensing_maps}
\end{eqnarray}
Here, $W_{\ell}^{G}$ and $W_{\ell}^{L}$ are the optimal linear filters \citep{hu07} to maximize the lensing S/N:
\begin{eqnarray}
W_{\ell}^{G} &=&   \left\{
\begin{array}{l l}
C^{\rm unl}_{\ell} (C_{\ell} + N_{\ell}^{\tszfreemapnotation})^{-1}&, ~\ell \le \ell_{\rm G}\\\notag
0&, ~{\rm otherwise}
\end{array}\right.\\
W_{\ell}^{L} &= & (C_{\ell} + N_{\ell}^{150})^{-1},
\label{eq_QE_filters}
\end{eqnarray} with $(C_{\ell}^{{\rm unl}})C_{\ell}$ corresponding to (un)lensed CMB temperature power spectra
calculated using the Code for Anisotropies in the Microwave Background \citep[\texttt{CAMB,}\footnote{\url{https://camb.info/}}][]{lewis00}.
$N_{\ell}$ is the noise spectrum for the indicated map,
after deconvolving the beam and filter transfer function given in Eq. (\ref{eq_filter_TF}).
We also add estimates of foreground power such as SZ, CIB, and radio galaxy emission, based on measurements by \citet{george15}, into $N_{\ell}$.
As described above, $\ell_{\rm G}$ is chosen to remove the magnification bias discussed by \citet{hu07} and additionally to suppress power from signals other than the primary unlensed CMB.
We set $\ell_G=2000$ for clusters with richness $\lambda < 60$.
For the rest, we use $\ell_G=1000$ as the convergence signal from these massive clusters can cause a negative bias in the estimate of the background gradient.
While this is a sharp change in $\ell_{\rm G}$, we will see below that it causes a negligible effect in our final \snr.

Although this method essentially eliminates the tSZ bias, creating a tSZ-free map can enhance other foregrounds (relative to the CMB) along with the noise. 
We 
look into possible biases from other foregrounds in \S\ref{subsec_tszbias} using the simulations from \citet[hereafter \citetalias{sehgal10}]{sehgal10}. 
 
\subsection{Simulations of the microwave sky}
\label{sec_sims_funda}
In this section, we describe the simulations used for the pipeline validation. 
We calculate the large-scale structure lensed CMB power spectra for the fiducial \planck\ 2015 cosmology \citep{planck15-13} using \texttt{CAMB}.
and create \boxsize\ Gaussian realizations of the CMB temperature map with $0.^{\prime}25$ pixel resolution.\footnote{We have confirmed that the results are unchanged when going to smaller initial pixels.}
Given the small angular extent, these simulations are done in the flat-sky approximation. 
These simulations are then lensed using the simulated galaxy cluster convergence profiles from the next section. 
Next we apply frequency-dependent foreground 
realizations (see \S\ref{sec_validation}). 
The simulated maps are convolved by the beam functions, and are rebinned to $0.^{\prime}5$ pixels to reduce the computational requirements. 

For realistic simulations, we must also account for the noise and the filtering applied to the real data. 
We add instrumental noise realizations corresponding to \sptpol{} maps (see \S\ref{sec_sptpol}).
We follow \citetalias{baxter18} and other SPT works and approximate the map filtering using a function of the form:
\begin{equation}
F_{\bar{\ell}} = e^{-(\ell_{1}/|\bar{\ell}|)^{6}}  e^{-(\ell_{2}/\ell_{x})^{6}} e^{-(\ell_{x}/\ell_{3})^{6}}.
\label{eq_filter_TF}
\end{equation}
We validate the robustness of this approximation in \S\ref{subsec_TF}.
For the small-scale lensing map, we set: \mbox{$\ell_{1}$ = 300}, \mbox{$\ell_{2}$ = 300}, and \mbox{$\ell_{3}$ = 20,000}. 
For the gradient map, we set: \mbox{$\ell_{1}$ = 0} (as the gradient map does not have an isotropic filter), \mbox{$\ell_{2}$ = 30}, and \mbox{$\ell_{3}$ = 3000}.


\subsection{Cluster convergence profile}
\label{sec_clus_profile} 
Now we summarize the method to model the convergence signal at cluster locations.
The total convergence \kappatotalmz{} profile for a galaxy cluster includes contributions from its own matter over-density (the {\it 1-halo} term) as well as from correlated structures along the line of sight (the {\it 2-halo} term; \citealt{seljak00a,cooray02}). 
For the {\it 1-halo} term, \kappaonehalomz, we use the Navarro-Frenk-White (NFW, \citealt{navarro96}) profile in Eq. (\ref{eq_nfw_density_with_delta_c}) to model the underlying dark matter (DM) density profile of the DES RM galaxy clusters: 
\begin{eqnarray} 
\rho\left(r\right) & = & \frac{\rho_0}{\left(\frac{r}{R_{\rm s}}\right)\ \left(1+\frac{r}{R_{\rm s}}\right)^2},
\label{eq_nfw_density_with_delta_c}
\end{eqnarray} 
where $R_{\rm s}$ is the scale radius, and $\rho_{0}$ is the central cluster density.
In \S\ref{subsec_clusprofile} we 
quantify the robustness of the inferred masses to this assumption by instead using the Einasto DM profile \citep{einasto89}. 
We use the photometric redshift measurements in the DES RM cluster catalog and use the \citet{duffy08} halo concentration formula to obtain the concentration parameter $c_{200}(M,z) = R_{200}/R_{\rm s}$. 
The convergence profile \kappaonehalofull\ at a radial distance $\theta$ for a spherically symmetric lens like NFW is the ratio of the surface mass density of the cluster and the critical surface density of the universe at the cluster redshift $\Sigma(\theta)/\Sigma(crit)$. 
To get the NFW convergence profile, we adopt the closed-form expression given by Eq. (2.8) of \citet{bartelmann96}. 

When evaluating the pipeline using mock cluster datasets we leave out the {\it 2-halo} term.
For the real data, we also consider the lensing arising from structures surrounding the cluster.
We model the {\it 2-halo} term contribution, \kappatwohalomz, to the total lensing convergence using Eq. (13) of \citet{oguri11}. 
The bias $b_h(M,z)$ of a halo with mass M $\equiv$ \mvir{} was calculated adopting the \citet{tinker10} formalism.
Finally, we correct the \refresponse{cluster} convergence profile \refresponse{\kappaonehalomz} for the uncertainties in the DES cluster centroids in \S\ref{sec_nfw_model_fitting}.


\subsection{Cluster cutouts}
\label{sec_cluster_cutouts}
We now describe the process of extracting cluster cutouts from \sptpol{} maps.
The lensing quadratic estimator described above is applied to these cutouts to reconstruct the lensing signal.
We extract \boxsize\ cutouts from the \sptpol{} temperature (tSZ-free and 150 GHz) maps around each cluster from the DES RM cluster catalog. 
This corresponds to a roughly $\sim$135 Mpc region around a cluster at $\tilde{z} = 0.77$. 
While the cutout size is much larger than the virial radius of the cluster, we emphasize it is necessary to robustly reconstruct the lensing signal using the background CMB.
This is because the amplitude of the lensing signal is proportional to the level of the background gradient, and the CMB has power on scales much larger than the typical cluster size of a few arcminutes. 
Performing the analysis with smaller cutouts will reduce the \snr{} of the estimated CMB gradient and affect the final lensing \snr.
After extracting the lensing signal, we limit the modeling and likelihood calculations to a $10^{\prime}$ region around the cluster.


\subsection{Stacked convergence and the weighting scheme}
\label{subsec_weights}
The lensing \snr{} for a single cluster is much less than unity, and we must stack the lensing signal from several clusters to achieve a reasonable \snr.
Thus the stacked convergence map is simply: 
 \begin{equation}
 \hat{\kappa} = \frac{\sum_j  w_j \left[ \hat{\kappa}_{j} - \left< \hat{\kappa}_{j} \right> \right]}{\sum_j w_j} - \hat{\kappa}_{\rm MF},
 \end{equation}
 where $\hat{\kappa}_{j}$ refers to reconstructed convergence map of cluster $j$ 
 and the weighting scheme $w$ is described below.
From the stacked map, we remove all modes above the \sptpol{} 150 GHz beam scale of $\theta_{\rm FWHM}\sim 1.^{\prime}2$. 
We also remove an estimate of the mean-field $\hat{\kappa}_{\rm MF}$ from this stacked convergence map.
The mean-field arises because of two reasons.
One because the temperature maps, before being filtered using Eq. (\ref{eq_QE_filtered_gradient_lensing_maps}), are apodized using a Hanning window\footnote{\url{http://mathworld.wolfram.com/HanningFunction.html}} with a $10^{\prime}$ edge taper to reduce edge effects.
The other reason is the presence of inhomogeneous noise in the survey region.
We obtain the mean field bias by stacking the convergence maps reconstructed at \howmanyrandomsforMF{} random locations in the maps.  

{\it Weighting scheme:} We decompose the weights for each cluster into two components: 
The first is the inverse-noise-variance weight, $w_{k}$, constructed from the observed standard deviation $\sigma_{\kappa}$ in the reconstructed \sptpol{} convergence maps in a ring between $10^{\prime}$ and $30^{\prime}$ around the cluster. 
The noise in convergence is proportional to the noise in the associated gradient map and increases, as expected, when $\ell_{\rm G}$ is reduced.
The second\footnote{We note that for the mean-field reconstructed from random locations, we only apply the weight $w = 1/\sigma_{\kappa}^2$ for stacking.} weight comes from the noise in the convergence maps due to the presence of tSZ signal in the second leg of the QE, the \sptpol{} 150 GHz map.
While our method completely eliminates the tSZ-induced lensing bias, the presence of tSZ signal in the second map tends to increase the variance in the convergence maps. 
The noise is proportional to the tSZ brightness and, as expected, is higher for massive clusters.
For example, the lensing signal of a cluster is proportional to its mass $M$ while the tSZ signal scales roughly as $M^{5/3}$.
\refresponse{We note that for the mean-field reconstructed from random locations, we only apply the weight $w = 1/\sigma_{\kappa}^2$ for stacking.} 

We obtain this second set of weights, $w_{\rm SZ}$, using simulations. 
For every cluster in the DES sample, we reconstruct the convergence profile using a simulated tSZ-free gradient map and a 150 GHz map with tSZ signal assuming an Arnaud profile \citep{arnaud10} with a log-normal scatter of 20\% in the $Y_{SZ}-M$ relation. 
We turn off cluster lensing, as the objective here is to only get an estimate of the tSZ-induced noise in the convergence maps.
A total of 25 simulations were used to get the noise estimate for each DES cluster.
The weights are estimated as
$w_{\rm SZ} = 1/\sigma_{\rm SZ}^{2}$, where $\sigma_{\rm SZ}$ is the standard deviation of the `null' convergence map within an angular distance of $10^{\prime}$ from the cluster centre.
The errors increase with richness and take a power law form parameterized as $\sigma_{\rm SZ}(\lambda) =  \sigma_{0}\lambda^{\alpha}$ with values $
(\sigma_{0}, \alpha) = (0.0045, 1.55)$.
The results are unchanged if we derive the weights using the tSZ signal from \citetalias{sehgal10}.
The total weight is now:
 \begin{equation}
 w = \frac{1}{\sigma_{\kappa}^{2} + \sigma_{\rm SZ}^{2}}.
\label{eq_cluster_weight}
 \end{equation}

Introducing $w_{\rm SZ}$ down-weights the most massive clusters, reducing the contribution
of clusters with $\lambda \ge 60$ to less than 1\% in the final stacked sample. 
This is why the change in gradient-map LPF scale to \mbox{$\ell_{\rm G} = 1000$} from the fiducial \mbox{$\ell_{\rm G} = 2000$}
for these clusters (see \S\ref{subsec_QE_modificaltion}) has negligible effects in our final results. 

An alternative to this down-weighting is to swap the maps in the two legs of the QE (i.e. the 150 GHz map for the gradient estimation and the tSZ-free map to reconstruct lensing) for clusters with \mbox{$\sigma_{\rm SZ} > \sigma_{\kappa}$}, which is approximately true for clusters with $\lambda > 40$. However, this results in a minimal gain as the \sptpol{} tSZ-free map has a higher noise ($\times3$) compared to the \sptpol{} 150 GHz maps. 
Some other approaches to handle the additional noise from the tSZ signal include (a) rotating the reconstructed lensing map based on the direction of the background CMB gradient and fitting for the tSZ-noise, (b) removing a matched-filter estimate of the tSZ-signal from the 150 GHz map before passing the map into the QE. We will explore such possibilities in detail in a future work (Patil S. et al. 2019, in preparation).
 
\subsection{Model fitting}
\label{sec_nfw_model_fitting}
We radially bin the stacked convergence map $\hat{\kappa}$ for the likelihood calculation in Eq. (\ref{eq_QE_likelihood}). 
To obtain the average lensing mass of the DES RM cluster sample, we need to compare this observed, radially binned, stacked convergence profile to convergence models generated using an assumed halo profile.
Essentially, we create a convergence model for every cluster using the NFW profile (see \S\ref{sec_clus_profile}) as a function of mass and the cluster redshift \kappaonehalomz, add the two halo term \kappatwohalomz, filter the model as per the real data, and then stack all the clusters using the weights described in the previous section.

The convergence model \refresponse{\kappaonehalomz} must be slightly modified to account for the uncertainties in the RM cluster centroids.
\citet{rykoff16} compared the centroids of DES RM clusters with SZ \citep{bleem15} and X-ray observations and found a fraction, $f_{\rm mis} = 0.22 \pm 0.11$, of the DES clusters to be mis-centered by $\sigma_{R}$, which is a fraction of the cluster radius $R_{\lambda} = (\lambda/100)^{0.2} h^{-1}$ Mpc. They further modelled the mis-centering as a Rayleigh distribution with $\sigma_{\rm R}  = c_{\rm mis} R_{\lambda}$ where ln $c_{\rm mis} = -1.13 \pm 0.22$.
Mis-centering ought to smear the convergence profiles and we use the prescription provided in Eq. (34) of \citet{oguri10} to account for the cluster mis-centering. We set \mbox{$f_{\rm cen} = 1- f_{\rm mis} =  0.78$} and $\sigma_{\rm s} = \sigma_{\rm R}/D_{\rm A}(z)$, where $\sigma_{\rm R}$ is picked from the Rayleigh distribution \citep{rykoff16}, and $D_{\rm A}(z)$ is the angular diameter distance at the cluster redshift $z$.
After the mis-centering correction, we filter the model using the approximation to the data filtering in Eq. (\ref{eq_filter_TF}) and remove all modes above the \sptpol{} 150 GHz beam similar to the data. 
The filtered model of all the individual clusters is weighted (\S\ref{subsec_weights}), stacked, and radially binned. 

With the model prediction in hand, we can then write down the likelihood of observing the real data as: 
\begin{multline}
-2\ln\mathcal{L (\hat{\kappa}(\theta)| \rm{M}}) = \\
\sum_{\theta = 0}^{10^{\prime}}\left[\hat{\kappa}(\theta) - \kappa(\theta)\right] \hat{{\rm C}}^{-1} \left[\hat{\kappa}(\theta) - \kappa(\theta)\right]^{T},
\label{eq_QE_likelihood}
\end{multline}
where $\hat{\kappa}(\theta),\ \kappa(\theta)$ are the azimuthally averaged radial profiles of the stacked data and model convergences, respectively, binned in 10 linearly spaced intervals with $\Delta\theta = 1^{\prime}$. 
To obtain the covariance matrix we use a jackknife re-sampling technique. We divide the \sptpol{} 500 \sqdeg\ region into $N=500$ sub-fields and estimate the covariance matrix for the radially binned convergence profile as 
\begin{equation}
\hat{{\rm C}} = \frac{N-1}{N} \sum\limits_{j=1}^{N = 500} \left[\hat{\kappa}_{j}(\theta) - \left<\hat{\kappa}(\theta)\right>\right] \left[\hat{\kappa}_{j}(\theta) - \left<\hat{\kappa}(\theta)\right>\right]^{T},
\label{eq_JK_cov}
\end{equation}
where $\hat{\kappa}_{j}(\theta)$ is the azimuthally binned stacked convergence of all the clusters in the $j^{th}$ sub-field and  $\left<\hat{\kappa}(\theta)\right>$ is the ensemble average of all the 500 sub-fields.
We test this approach by alternatively estimating the covariance matrix using 500 realizations of the random convergence stacks. 
We do not note any significant differences between the uncertainties estimated using the two approaches. 
We apply the \citet{hartlap06} correction to $\hat{{\rm C}}^{-1}$ to account for the noise in covariance estimation due to the finite number of jackknife re-sampling.



\section{Data and pipeline validation}\label{sec_validation}
In this section, we describe tests used to investigate the known and unknown systematic effects in the data and to validate the pipeline.
We start with the test for unknown systematics through the ``curl'' null test (\S\ref{sec_null_tests}).
Next we calculate the expected systematic error budget from known sources of systematic uncertainty (\S\ref{sec_sys_checks}).

\subsection{``Curl'' null test}\label{sec_null_tests}

\begin{figure}
\centering
\ifdefined\ApJsubmit
\includegraphics[width=0.45\textwidth, keepaspectratio]{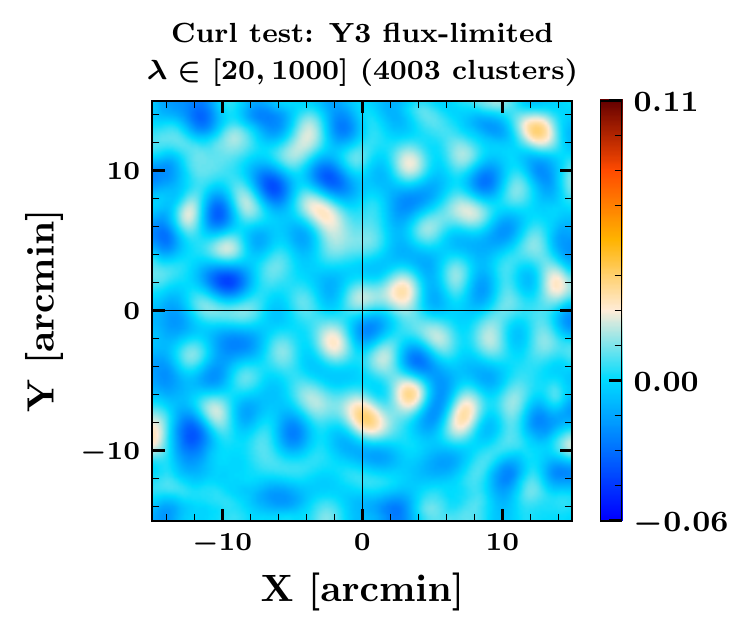}
\else
\includegraphics[width=0.45\textwidth, keepaspectratio]{figs/kappa_model_MF_y3_v6_4_22_full_curl_test_JODY.pdf}
\fi
\caption{Stacked result of the curl test performed at the cluster locations by replacing the divergence operator in Eq. (\ref{eq_QE_kappa}) with a curl operator. 
We obtain a PTE value of 0.26, consistent with a null result.
For the ease of visual comparison we adopt the same colour scale as in Fig. \ref{fig_QE_stacked_maps}.
}
\label{fig_QE_null_tests}
\end{figure}

We perform a ``curl'' null test \citep{hu07} at \howmanyclustersinfullsample{} cluster locations from the DES RM \whichyear{} flux-limited sample.
Specifically, we replace the divergence of the gradient field, $\nabla \cdot [G(\bnhat) L^{*}(\bnhat)]$, in Eq. (\ref{eq_QE_kappa}) with the curl operator. 
Since the curl of a gradient field is zero, the reconstructed field should be consistent with zero unless there is a systematic bias in the data. 
The result of the curl test is shown in Fig. \ref{fig_QE_null_tests}. 
We radially bin the test result similar to the cluster stack as described in \S\ref{sec_nfw_model_fitting} and compare it to a zero signal.
The test returns a probability to exceed (PTE) value of 0.26, consistent with a null signal.

\subsection{Systematic error budget}\label{sec_sys_checks}

Now we consider possible sources of systematic error. 
 We estimate the bias due to each cluster's tSZ emission and residual foregrounds, the assumption of an underlying cluster profile, uncertainties in the DES RM mis-centering parameter $f_{\rm mis}$, approximations to the filter transfer function (Eq.~\ref{eq_filter_TF}), uncertainties in the beam measurements, and the assumption of a background cosmology.
Another source of systematic error is the uncertainties in the cluster redshifts estimated photometrically.
However impact of photo$-z$ errors was estimated to be negligible by \citetalias{raghunathan17a}, and we ignore them here.

We rely on the \citetalias{sehgal10} simulations to estimate the level of residual-tSZ/foreground bias in the RM \whichyear{} flux-limited sample.
In all the other cases we use the data and report the shift in the average lensing mass of the clusters in the RM \whichyear{} volume-limited sample obtained in \S\ref{sec_results}. 
The combined systematic error budget is presented in Table~\ref{tab:sys}.
The systematic error \refresponse{calculated as a quadrature sum of the errors presented in Table~\ref{tab:sys}} is much smaller than the statistical error in the measurements at a level of \refresponse{$0.15\sigma$}. 
\refresponse{Using a direct sum, the combined error budget is $0.27\sigma$.}
The dominant error comes from the uncertainty in the DES RM cluster centroids shifting the mean lensing mass by 2.8\%. 

\begin{table}
\caption{Systematic error budget \refresponse{in the stacked mass} for DES RM \whichyear{} volume-limited sample}
\footnotesize{
\centering
\begin{tabular}{| l | C{1.7cm}| C{2cm} |}
\hline
\multirow{2}{*}{Source of error} & \multicolumn{2}{c|}{Magnitude of error}\\
\cline{2-3}
& \%  & frac. of $\sigma_{stat}$ \\\hline
Beam uncertainties & $<0.01$\% & - \\\hline
Cluster mis-centering & 2.78\% & $0.12\sigma$ \\\hline
Cosmology & $0.39\%$ & $0.03\sigma$ \\\hline
Filtering $\ell_{x}$ &  0.21\% & 0.02$\sigma$\\\hline
Halo profile &  0.12\% &  0.01$\sigma$ \\\hline\hline
\refresponse{Residual foregrounds} & \refresponse{2.12\%} & \refresponse{$0.09\sigma$}\\\hline
Total & \refresponse{3.53\%} & \refresponse{0.15$\sigma$} \\\hline
\end{tabular}
}
\label{tab:sys}
\tablecomments{
This is a list of systematic errors estimated for the lensing mass measurement.
}
\end{table}

\subsubsection{Cluster tSZ signal and residual foregrounds}\label{subsec_tszbias}

\begin{figure}
\centering
\ifdefined\ApJsubmit
\includegraphics[width=0.45\textwidth, keepaspectratio]{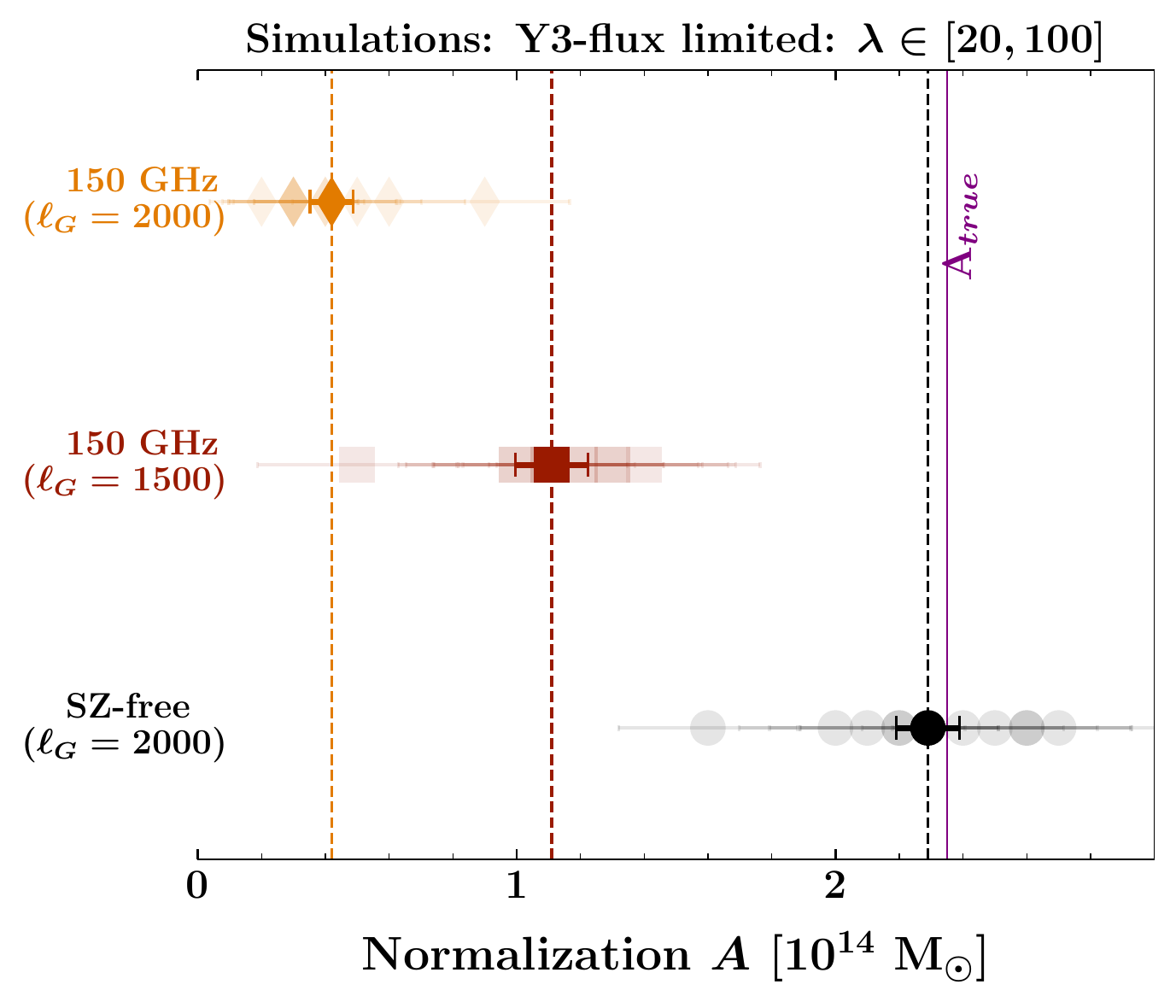}
\else
\includegraphics[width=0.45\textwidth, keepaspectratio]{figs/tsz_bias_checks_sehgal_sims.pdf}
\fi
\caption{Quantifying the level of bias due to residual foregrounds and the tSZ signal using \citetalias{sehgal10} simulations. The recovered lensing mass, un-biased for the fiducial case with \tszfreemapnotation\ map for gradient estimation and $\ell_{\rm G}$ = 2000 is shown as black circles. The equivalent biased results with just the 150 GHz map and $\ell_{\rm G}$ = 1500 (2000) cutoff scales for the gradient estimation are shown as red squares (orange diamonds). Each light shaded point corresponds to an individual simulation run with clusters from the DES RM \whichyear\ flux-limited sample. The darker data points are the values obtained for $10\times$ the sample size.}
\label{fig_QE_sehgal_sims}
\end{figure}

In this work we eliminate the bias due to tSZ signal in the reconstructed lensing maps using \tszfreemapnotation\ maps to estimate the background gradient of the CMB. 
However, projecting just the tSZ signal out tends to modify other frequency-dependent foregrounds, and the resultant map is not an optimal foreground-free CMB map for the lensing reconstruction. 
This enhancement of foregrounds generally acts as an additional source of noise and tends to increase the variance of the reconstructed lensing maps. 
At the cluster locations, however, an increase in foreground emission due to galaxies inside the cluster can introduce undesired mode coupling between the estimated gradient map and the lensing map resulting in a biased lensing signal. 
Since massive clusters host more galaxies, we can expect the bias to increase with the cluster mass or equivalently richness. 
Here we quantify this bias using the \citetalias{sehgal10} foreground simulations.\footnote{\url{https://lambda.gsfc.nasa.gov/toolbox/tb_sim_ov.cfm}}

To this end, we begin with the simulated skies described in \S\ref{sec_sims_funda}, to which we then add simulated clusters, including the lensing signal (only the {\it 1-halo} term), thermal and kinetic SZ effects, and emission associated with the cluster (e.g. from member galaxies). 
These simulations also include foregrounds uncorrelated with the clusters such as field radio galaxies.
The addition of foregrounds using \citetalias{sehgal10} simulations is described below. 
Note that the foreground maps, whether associated with the cluster or not, are not lensed by the cluster in these simulations. 
The number of simulated clusters and their redshifts and richnesses are derived from the  DES RM \whichyear{} flux-limited sample. 
The richnesses and redshifts are converted to cluster masses according to the \ML{} relation (Eq.~\ref{eq_ML}) with best-fit parameters from \citet{melchoir17}, $A_{\rm M17} = 2.35\ \times$ \munits, $\alpha_{\rm M17} = 1.12$, and $\beta_{M17} = 0.18$.  

For foregrounds, we extract half-arcminute resolution \boxsize\ cutouts of the 95 and 150 GHz \citetalias{sehgal10} simulations of the tSZ, kSZ, radio, and infrared galaxies around halos corresponding to the mock cluster sample. 
We scale the tSZ power down from the \citetalias{sehgal10} simulations by a factor of 1.75 to match the \citet{george15} measurements.
The \citetalias{sehgal10} simulations contain $\sim$16,000 halos above \mvir\ $\ge$ \munits\ and 175 halos above \mvir\ $\ge 5 \times$ \munits\ at redshifts $z \ge 0.25$.
These foreground cutouts are added to our mock galaxy cluster lensed CMB datasets. 
The maps are then processed in the same way as explained in \S\ref{sec_method_QE} to extract the tSZ cleaned map and passed into the QE. 


We present the results in Fig. \ref{fig_QE_sehgal_sims}.
The true normalization is shown as the purple solid line.
In the figure, the light shaded data points are the result for a single simulation run ($\sim$ 4000 clusters) and the darker data points are the results for $10\times$ the sample size. 
For our baseline analysis with \tszfreemapnotation{} maps for the gradient estimation and $\ell_{\rm G} = 2000$, the recovered normalization is $\la0.5\sigma$ from the true value.
We obtain \mbox{$A = 2.30 \pm 0.09\ \times$ \munits} (black circle) implying no significant residual foreground bias in the lensing measurements.
This result also provides evidence that the lensing pipeline is unbiased.

{\it Comparison to standard QE:} We also use the \citetalias{sehgal10} simulations to compare the modified QE to the standard case using the 150 GHz map for the background CMB gradient estimation.
In the standard case, the correlations introduced between the two maps by the foregrounds, the tSZ signal in particular, can be alleviated by lowering the LPF threshold $\ell_{\rm G}$ for the gradient map as in Eq. (\ref{eq_QE_filters}). 
As described in \S\ref{sec_method_QE}, the choice of $\ell_{\rm G}$ is a trade-off between the level of foreground bias and the lensing \snr{}. 
Here we adopt $\ell_{\rm G} = 2000$ and $\ell_{\rm G} = 1500$ and note that the results are heavily biased in both cases: red squares (orange diamonds) for $\ell_{\rm G} = 1500$ ($\ell_{\rm G} = 2000$).
The level of bias is higher when $\ell_{\rm G}$ is set to 2000 compared to 1500, as expected.
This bias is predominantly due to the tSZ signal and can be reduced by removing massive clusters from the analysis as in \citetalias{baxter18}. 
For comparison, when we apply a richness cut of $\lambda \in [20, 40]$ the lensing bias is reduced from 82\% to 65\% for $\ell_{\rm G} = 2000$ and 52\% to 35\% for $\ell_{\rm G} = 1500$. 
This cut removes $\sim$500 massive clusters from the analysis.
This result can be compared to the conservative tSZ-bias of 11\% set by \citetalias{baxter18} with $\ell_{\rm G} = 1500$ for the same richness range $\lambda \in [20, 40]$.
\citetalias{baxter18} obtained a lower bias value as the high-$\ell$ modes in the SPT-SZ maps are down-weighted due to $4\times$ higher noise.
This also suggests that we cannot handle the tSZ bias by simply removing clusters above a certain richness, for example $\lambda \ga 40$, for low-noise CMB datasets.

Finally, a subtle point from the figure is that the mass constraints obtained using the 150 GHz map for gradient estimation
(orange diamonds) are better ($\sim$ 14\%) than those obtained using the tSZ-free map for gradient estimation (black circles) despite adopting $\ell_{\rm G} = 2000$ in both cases.
This hit in the \snr{} arises because the \sptpol{} \tszfreemapnotation{} map is noisier than the 150 GHz.

\subsubsection{Cluster profile}\label{subsec_clusprofile}
In our fiducial analysis, we assume that the underlying halo profile of the clusters follows the NFW dark matter model. 
However,halos in real clusters 
deviations from the NFW profile have been observed \citep[e.g.,][]{diemer14}, and 
\citet{child18} argued that the Einasto model is a better fit than NFW to stacked halo profiles.

In this section, we estimate the magnitude of a possible bias due to the assumption of the incorrect mass profile by using an Einasto profile \citep{einasto89} to model the lensing convergence $\kappa^{1h}$.
The lensing QE and subsequently the reconstructed convergence maps remain unchanged.
The Einasto profile is defined as
\begin{eqnarray}
\rho(r)_{_{\rm Ein}} & = &  \rho_{_{0}}\ \mbox{exp}\left( - \frac{2}{\alpha} \left[\left(\frac{r}{R_{\rm s}} \right)^{\alpha} - 1\right]\right),
\label{eq_einasto_density}
\end{eqnarray}\\
where $\alpha = 0.18$ is the shape parameter \citep{ludlow13}. 
As in the NFW analysis, the concentration $c_{200}$ as a function of mass and cluster redshift is obtained using the \citet{duffy08} relation.
We use the general framework for spherically symmetric halos defined in \citetalias{raghunathan17a} and simply plug the above density profile into Eq. (2.9) of \citetalias{raghunathan17a} to obtain the Einasto convergence \kappaonehalo$^{,\rm Ein}$ profiles. The \kappatwohalo\ term remains the same. For the Einasto case we see a negligible shift of $0.01\sigma$ compared to our fiducial result.


\subsubsection{Uncertainties in filter transfer function and beam}\label{subsec_TF}
As described in \S\ref{sec_data}, the \sptpol{} map-making process is lossy, with noisy modes along the scan direction filtered out. 
The ideal, if computationally expensive, approach to handle the filtering would be an end-to-end simulation from the TOD to the lensing reconstruction. 
In this work, we take a computationally much cheaper approach and approximate the filtering by the phenomenological fit to the filter transfer function in Fourier space given by Eq. (\ref{eq_filter_TF}). 
The major uncertainty is in the position of the high-pass filter (HPF) in the scan direction: this filters modes more strongly than the isotropic HPF, and the LPF is at angular scales that do not matter to the reconstruction. 
The estimated position for this HPF is $\ell_x = 300 \pm 20$. 
We also recompute the models for an assumed $\ell_x = 280$ and 320 to evaluate the shifts in the lensing masses.
We note no significant effect (masses shift by roughly $\pm 0.02\sigma$), indicating that the uncertainty in the simplified filtering treatment causes negligible changes to our results.

Similarly, we also check the effect of errors in the telescope beam modeling \beambl{} that were derived using Venus observations (see \S\ref{sec_sptpol}). 
We find that the effect due to beam uncertainties in the final result is also negligible. 
The shift in the lensing mass is $\la 0.01\sigma$ when we modify \beambl{} $\rightarrow$ \beambl{} $+\ 2\sigma$.

\begin{figure*}[t!]
\centering
\ifdefined\ApJsubmit
\includegraphics[width=1.\textwidth, keepaspectratio]{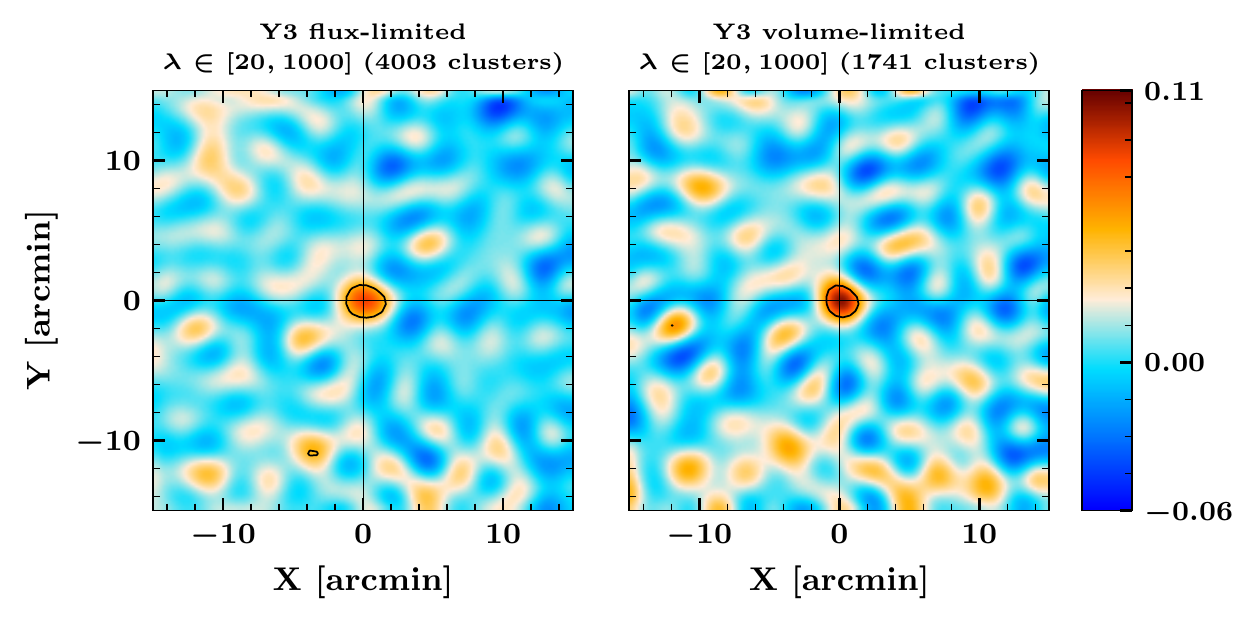}
\else
\includegraphics[width=1.\textwidth, keepaspectratio]{figs/kappa_model_MF_y3_v6_4_22_full_vl_JODY.pdf}
\fi
\caption{The inverse-variance weighted stacked convergence maps at the location of \howmanyclustersinfullsample\ and \howmanyclustersincosmosample\ clusters in the range $\lambda \in [20,1000]$ from the flux-limited (left) and the volume-limited (right) samples of the DES RM \whichyear\ cluster catalog.
The contour corresponds to the regions above $3.5\sigma$. 
The null hypothesis of no-lensing is rejected at \howmanysigmaforfullsample{} and \howmanysigmaforcosmosample{} for the two cases respectively.
}
\label{fig_QE_stacked_maps}
\end{figure*}

\begin{figure*}
\ifdefined\ApJsubmit
\includegraphics[width=1.\textwidth, keepaspectratio]{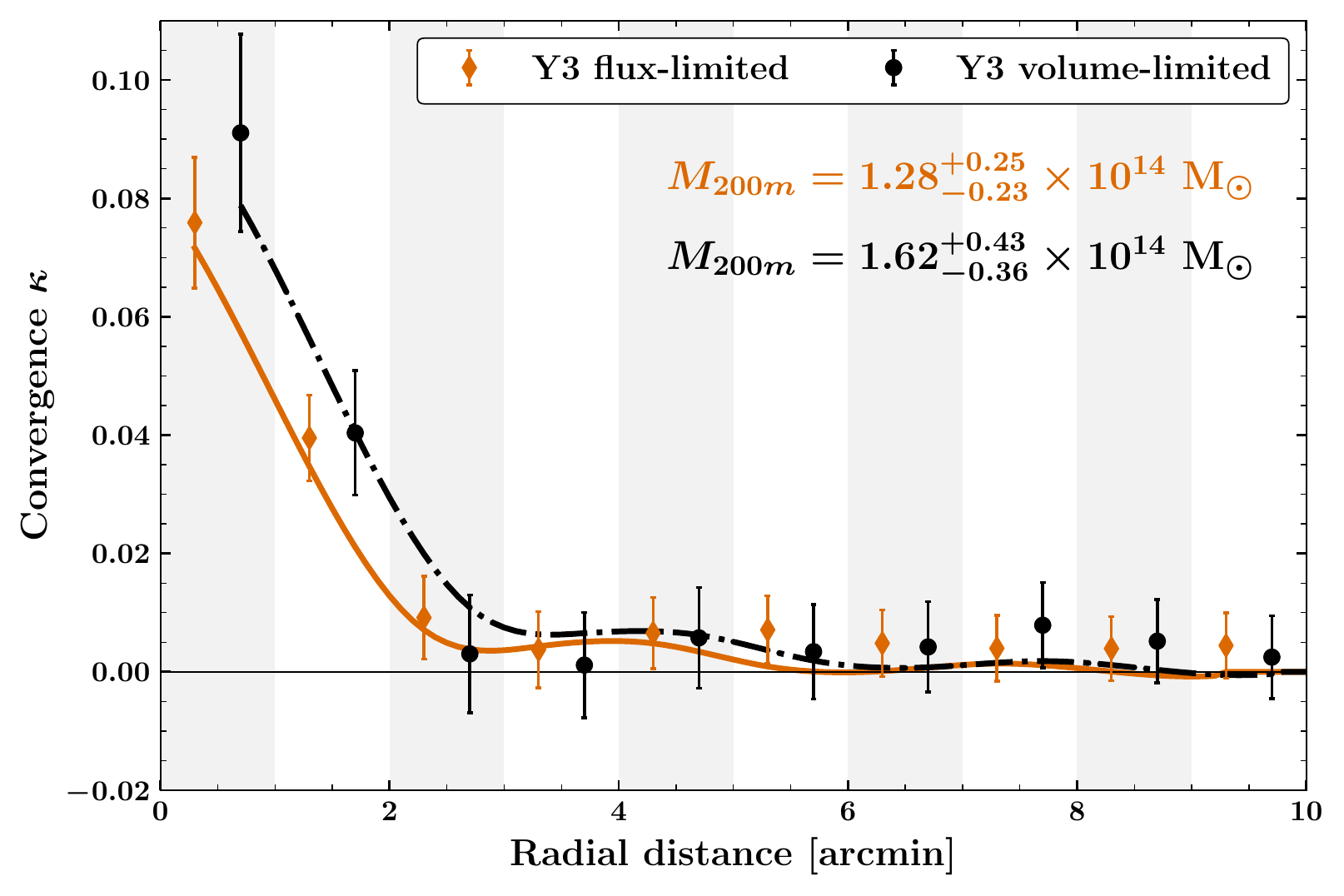}
\else
\includegraphics[width=1.\textwidth, keepaspectratio]{figs/kappa_model_MF_y3_v6_4_22_full_vl_radprf_JODY.pdf}
\fi
\caption{The azimuthally averaged radial profile of the stacked convergence maps from Fig. \ref{fig_QE_stacked_maps}. The black circles and orange diamonds correspond to the flux and volume-limited \whichyear{} DES RM cluster samples. The error bars are the diagonal value of the covariance matrix estimated using the jackknife technique in Eq. (\ref{eq_JK_cov}). The data points for the two samples have been artificially shifted from the bin centres to avoid cluttering.}
\label{fig_QE_stacked_maps_radprf}
\end{figure*}

\subsubsection{Uncertainties in $f_{\rm mis}$ parameter}\label{subsec_mis-centering}
In our baseline analysis we perform a mis-centering correction of the cluster convergence models (see \S\ref{sec_nfw_model_fitting}) using $f_{\rm mis} = 0.22$ based on the results by \citet{rykoff16} for the RM clusters from the science verification data.
Now we generate new convergence models assuming a larger fraction, 33\%, of the clusters are mis-centered by modifying the mis-centering parameter by its $1\sigma$ error 
from \citet{rykoff16}. 
Since the two parameters, $f_{\rm mis}$ and $c_{\rm mis}$, describing the cluster mis-centering are highly correlated, we also modify \mbox{ln $c_{\rm mis}$} = -1.32 for this test. 

The recovered mass increases by 2.8\% in this case.
However, the shift is only $0.12\sigma$, $1/8^{th}$ of the statistical uncertainty. 
The direction of the shift is consistent with expectations, as assuming a larger $f_{\rm mis}$ should smear the convergence model more than the fiducial case leading to an increased lensing mass. 
Since this is the dominant systematic, we also estimate the error for the flux-limited sample.
The mean lensing mass of the volume-limited sample goes up 3.2\% but the change is still smaller than the statistical error ($0.16\sigma$). 

\subsubsection{Underlying cosmology}\label{subsec_others}
The systematic error arising due to the assumption of a background cosmology is quantified here.
As described in earlier sections, in our fiducial analysis we use the \lcdm\ cosmology obtained using the \planck\ 2015 datasets \citep{planck15-13}. 
Here we repeat the analysis by modifying the lensed CMB power spectra $C_{\ell}$ to include the $1\sigma$ errors to the \planck\ 2015 cosmological parameters.
Modifying the background cosmology alters the weights of Eq. (\ref{eq_QE_filters}) in the lensing estimator and also the model convergence profiles \kappaonehalomz\ and \kappatwohalomz. 
However, the effect due to background cosmology in the inferred lensing mass is negligible with a shift in the lensing mass $\la 0.03\sigma$.

\section{Results and Discussion}\label{sec_results}

The main results of this work are the lensing-derived cluster mass constraints for the DES RM \whichyear\ cluster samples using \sptpol{} tSZ-free $\times$ 150 GHz temperature maps.
Below, we first present the lensing mass estimates in \S\ref{sec_temp_results} and use the lensing measurements from the DES \whichyear{} volume-limited sample to independently calibrate the \ML{} relation of the cluster sample in \S\ref{sec_ML_scaling}. 
Finally in \S\ref{subsec_lit_comparison} we compare our results to literature. 

\subsection{Stacked mass measurements}
\label{sec_temp_results}
In Fig. \ref{fig_QE_stacked_maps}, we present the results of our stacked lensing measurements. 
The left (right) panel correspond to the convergence maps stacked at the location of clusters in the DES \whichyear\ flux(volume)-limited sample.
The variance in the flux-limited sample is lower than the volume-limited sample because the flux-limited sample has twice as many objects. 
An estimate of the mean-field has been subtracted from the maps.
We reject the null hypothesis of no lensing with a significance of \howmanysigmaforfullsample\ for the flux-limited sample of \howmanyclustersinfullsample\ clusters. 
The obtained \snr{} is consistent with our expectations from the simulations shown as lighter black circles in Fig. \ref{fig_QE_sehgal_sims}.
For the smaller volume-limited sample, the no-lensing hypothesis is ruled out at  \howmanysigmaforcosmosample.

The radially binned convergence profiles that are used to estimate the cluster masses are shown in Fig. \ref{fig_QE_stacked_maps_radprf} along with the best-fit model curves. 
\refresponse{The PTE values for the best-fit convergence models are 0.68 and 0.65 for the full- and volume-limited samples respectively}.
The ringing pattern is because of the sharp filtering of modes above the \sptpol{} beam scale. 
The error bars plotted are the square root of the diagonal entries of the covariance matrix estimated using Eq. (\ref{eq_JK_cov}). 
As explained in \S\ref{sec_nfw_model_fitting}, all the mass estimates are derived by fitting a NFW profile along with the contribution from the {\it 2-halo} term to the measured radially binned profile. 
The recovered lensing masses for the stacked flux and volume-limited samples are \mbox{\mvir = \fitmassforfullsamplewitherrors\ $\times$ \munits} and \mbox{\fitmassforvlsamplewitherrors\ $\times$ \munits} respectively. 
According to expectations, the lensing masses shift up by 0.3$\sigma$ when the {\it 2-halo} term is excluded.

A higher mean mass is expected for the volume-limited sample. 
At redshifts above $z\sim 0.6$, galaxies at the luminosity threshold adopted by RM become too faint to be detected in the DES data. 
Consequently, the richness of the clusters is extrapolated from the subset of galaxies that are sufficiently bright to be detected. 
This extrapolation introduces additional noise in the richness estimates.
The increased scatter leads to more low-mass systems scattering up to apparently rich systems, thereby lowering the mean mass of the selected halos. 
For this reason, we restrict our analysis to the volume-limited sample in the subsequent sections. 

\subsubsection{Comparison to \citetalias{baxter18} analysis}
\label{sec_b18_corss_check}
Now we discuss the differences in the analysis choices between \citetalias{baxter18} and this work to compare the lensing \snr{}, $8.1\sigma$ vs. \howmanysigmaforcosmosample, obtained in the two works.\footnote{We perform the comparison with the volume-limited sample as \citetalias{baxter18} also performed the analysis with the DES Year-1 volume-limited sample}
\citetalias{baxter18} used the 2500 \sqdeg{} \sptsz{} data and DES RM Year-1 volume-limited sample in range $\lambda \in [20,40]$ with $2\times$ more\footnote{
Despite $5\times$ larger sky coverage, the \citetalias{baxter18} cluster sample was only $2\times$ larger than the one used here, because the overlap between the SPT-SZ survey and DES Year-1 data was only 40\%, compared to nearly full overlap between DES three-year data and the SPTpol survey.}
clusters than this work. 
Here we use the \sptpol{} 150 GHz map which is $\sim4\times$ deeper than the \sptsz{} survey. 
However, as shown in Fig. 2 of \citetalias{raghunathan17a}, the presence of foregrounds sets a floor to the achieved performance at low noise levels and we note that the improvement in \snr{} does not follow a simple scaling based on noise level.


\subsection{Mass-richness \ML{} scaling relation calibration}\label{sec_ML_scaling}

We now apply the lensing mass measurements from \S\ref{sec_temp_results} to constrain the relationship between a cluster's mass, $M$, and optical richness, $\lambda$, in the DES RM \whichyear{} volume-limited sample. 
We limit the analysis to just the volume-limited sample since the flux-limited sample has selection bias as explained above in \S\ref{sec_temp_results}.
Following earlier weak-lensing analyses of RM clusters \citep{simet18, melchoir17, mcclintock18}, we use a power-law scaling relation for cluster mass, $M$, as a function of richness, $\lambda$, and redshift, $z$,
\begin{eqnarray}
M = A \left(\frac{\lambda}{40}\right)^{\alpha} \left(\frac{1+z}{1+0.35}\right)^{\beta},
\label{eq_ML}
\end{eqnarray}
where $A$ is a normalization parameter, and the exponents $\alpha$ and $\beta$ are richness and redshift evolution parameters respectively. 
The pivot points for the richness and redshift evolution were set based on DES weak-lensing measurements of \citetalias{mcclintock18}.
The model for the stacked mass is 
\begin{equation}
M(A, \alpha, \beta) \equiv M = \frac{\sum_j w_j M(\lambda_{j}, z_{j})}{\sum_j w_j}, 
\label{eq_model_ML}
\end{equation}
where the sum runs over the number of clusters in the sample and the weight $w$ for each cluster is given in Eq. (\ref{eq_cluster_weight}).

\begin{figure}
\centering
\ifdefined\ApJsubmit
\includegraphics[width=0.45\textwidth, keepaspectratio]{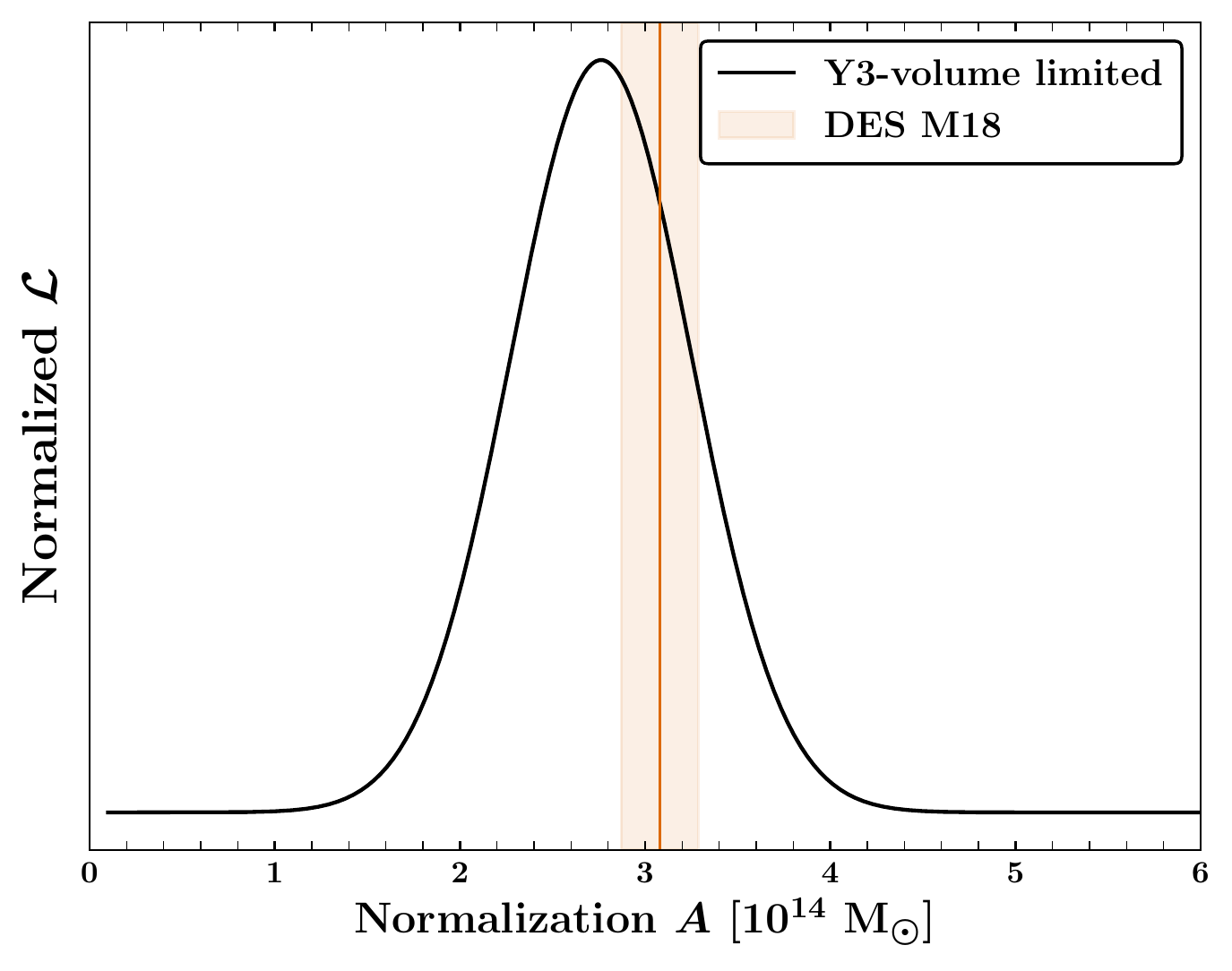}
\else
\includegraphics[width=0.45\textwidth, keepaspectratio]{figs/M_rich_fitting_y3_v6_4_22_full_vl_JODY.pdf}
\fi
\caption{Marginalized posteriors of the normalization parameter $A$ of the \ML\ relation for the volume-limited sample of the RM cluster catalog. 
The result is consistent with the best-fit values obtained by DES weak-lensing measurements \citepalias{mcclintock18}, shown as the shaded region.
} 
\label{fig_massrichness_fitting}
\end{figure}

We do not split the stacks into different richness or redshift bins.
As a result, the data's sensitivity to the two evolution parameters is minimal and we apply informative priors to both. 
We perform a Markov chain Monte Carlo (MCMC) analysis 
using the publicly available \emph{emcee} \citep{mackey13} code to sample the likelihood space.
We assume a flat prior for the normalization parameter $A$ in the range $A \in [0.1,100]\ \times$ \munits. For the slopes, we use a Gaussian prior based on \citetalias{mcclintock18}. Specifically, we set $(\alpha_{0}, \sigma_{\alpha})= (1.356, 0.056)$ and $(\beta_{0}, \sigma_{\beta})= (-0.3, 0.3)$. 
The posteriors on both the richness evolution parameters, as expected, follow the assumed prior. 
We obtain a normalization value of \mbox{$A$ = \fitAforvlsamplewitherrors $\times$ \munits{}}, which is consistent with results from DES weak-lensing measurements of $A_{\rm DES} = 3.08 \pm 0.21 \times$ \munits{} by \citetalias{mcclintock18}. 
The marginizalized posterior for $A$ is shown as the black curve in Fig. \ref{fig_massrichness_fitting} along with measurements from \citetalias{mcclintock18} as the orange shaded region.

\subsection{Comparison to literature}\label{subsec_lit_comparison}
\begin{figure}
\centering
\ifdefined\ApJsubmit
\includegraphics[width=0.48\textwidth, keepaspectratio]{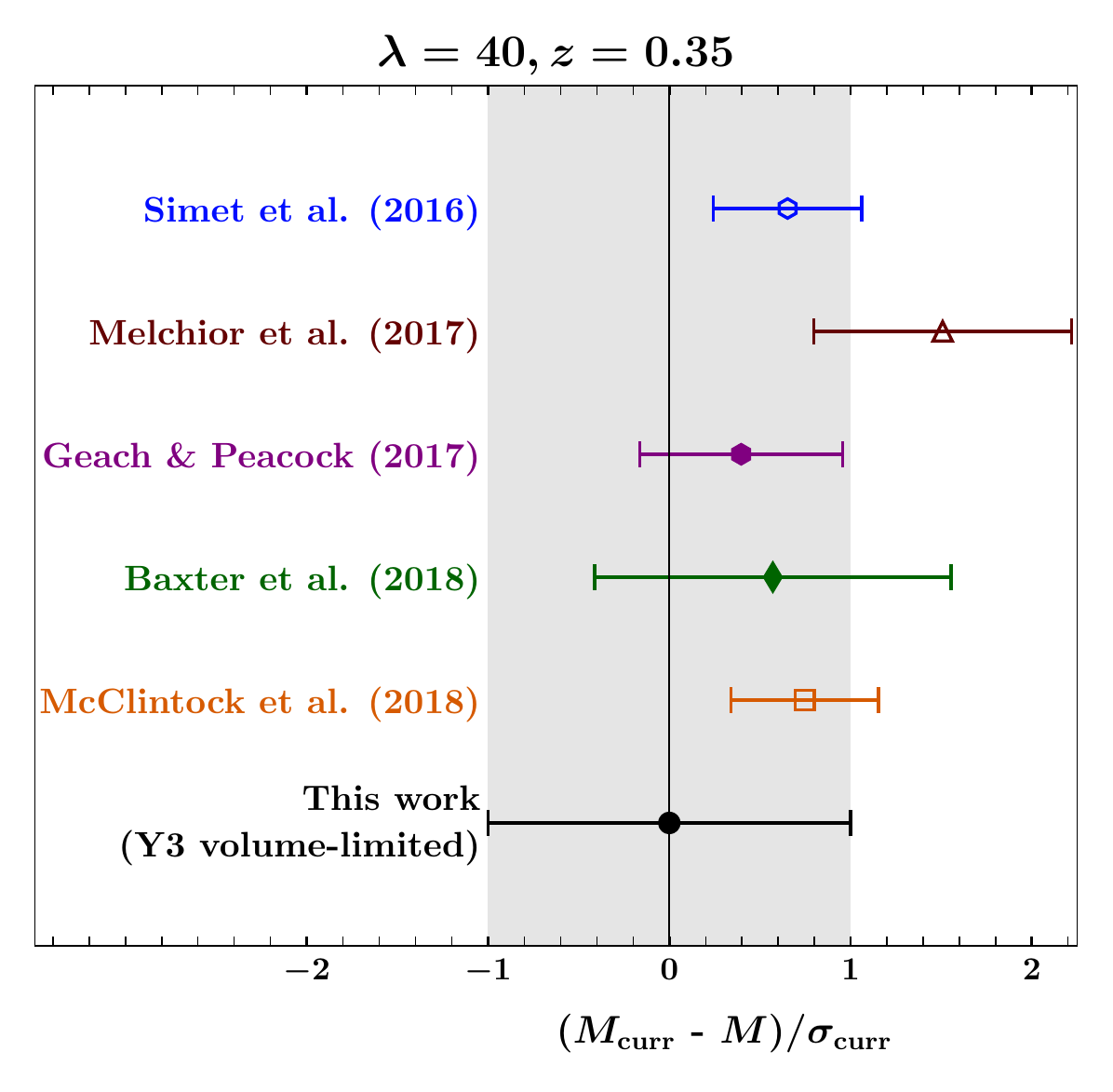}
\else
\includegraphics[width=0.48\textwidth, keepaspectratio]{figs/lit_comparisons_JODY.pdf}
\fi
\caption{Comparison of \mvir{} mass estimates of galaxy clusters obtained using the \ML{} relation from different works in the literature using the RM cluster catalogs. 
The points have been normalized using the $1\sigma$ error from the analysis with the \whichyear{} volume-limited sample of the DES RM catalog.
The filled (open) points represent measurements using the CMB-cluster lensing (galaxy weak-lensing) technique.
}
\label{fig_lit_comparison}
\end{figure}

We now compare our results to similar works from the literature performed with the RM cluster catalogs from the SDSS and DES experiments. 
Since the richness estimated for a given cluster from surveys A and B can vary slightly depending on the adopted data reduction and analysis choices, we include a a small correction factor $\epsilon_{\rm A{\text -}B}$ when comparing results from two surveys.
We compute the ratio $\lambda_{\rm A}/\lambda_{\rm B}$ for the overlapping clusters in the two surveys and simply set $\epsilon_{\rm A{\text -}B}$ to the median value of the ratios.
We find the richness estimates in DES Year-3 and Year-1 to be consistent with $\epsilon_{\rm Y3{\text -}Y1} = 1$. 
For the rest, we set: $\epsilon_{\rm Y3{\text -}SV} = 1.08$ and $\epsilon_{\rm Y3{\text -}SDSS} = 0.93$ \citepalias{mcclintock18}.
The comparison after including this correction factor is presented in Fig. \ref{fig_lit_comparison}, which is similar to Fig. 15 in \citetalias{mcclintock18}. 
Specifically, we show the difference in \mvir{} masses obtained from different works for a cluster with richness $\lambda = 40$ at redshift $z=0.35$, the pivot points in Eq. (\ref{eq_ML}).
The figure is normalized using the $1\sigma$ error from the current work with the \whichyear{} volume-limited DES RM catalog sample.

Each analysis uses a different cluster sample and lensing data.
\citet{simet18} and \citet{geach17} use the SDSS RM catalog sample containing roughly 26,000 clusters. 
\citet{melchoir17} use the full catalog from the DES science verification data while \citetalias{baxter18} and \citetalias{mcclintock18} perform the analysis using the DES Year-1 volume-limited sample.
The works by \citet{geach17} and \citetalias{baxter18} use the CMB-cluster lensing technique (filled points) with \planck{} and {\sc SPT-SZ} CMB temperature maps.
All the others use galaxy weak-lensing measurements and are represented as open points. As evident from the figure, our results are consistent with other similar works in the literature.

\section{Conclusions}\label{sec_conclusion}

We have built a modified lensing QE to reconstruct lensing potential at the location of the DES RM clusters using the \sptpol{} 500 \sqdeg{} field CMB temperature maps.
We detect a stacked lensing signal at \howmanysigmaforfullsample{} and \howmanysigmaforcosmosample{} level for the flux- and volume-limited samples of the \whichyear{} RM cluster catalog.
The modified QE eliminates the tSZ-induced lensing bias by using two maps for lensing reconstruction: a low-noise \sptpol{} 150 GHz map to reconstruct the small-scale lensing, and a tSZ-free map to estimate the background CMB gradient.
The tSZ-free map is internal, constructed from the \sptpol{} 95 and 150 GHz channels. 

We model the lensing signal, assuming a NFW profile for the galaxy clusters, and find the stacked lensing masses to be \mbox{\mvir = \fitmassforfullsamplewithsyserrors} and  \mbox{\fitmassforvlsamplewithsyserrors} $\times$ \munits{} for the two catalog samples. 
The uncertainties in our knowledge about the cluster centroids are the dominant contributor ($\sim3\%$) to the systematic error budget.
We use the mass measurements from the volume-limited sample to calibrate the mass-richness \ML{} scaling relation of the RM galaxy clusters.
The constraints on the richness and redshift evolution parameters are dominated by the priors assumed from the DES weak-lensing measurements \citepalias{mcclintock18}.
We obtain a best-fit normalization parameter of \mbox{$A$ = \fitAforvlsamplewitherrorsformatted $\times$ \munits}.
The results are consistent with other similar works in the literature performed using the RM catalogs from DES and SDSS surveys.
\refresponse{
It must be noted that one must account for the Malmquist and Eddington biases \citep[e.g.,][]{allen11} when using the above scaling relation parameters to infer individual cluster masses for cosmological analysis with cluster abundance measurements. 
The Malmquist bias is due to selection effects while the Eddington bias arises due to uncertainties in the inferred cluster masses which tends to up-scatter the low mass clusters into higher mass bins.
An an example, see the corrections performed by  \citet{battaglia16} for the ACT and \citet{pennalima17} for the {\it Planck} cluster samples respectively.
However, these corrections are not required for the current analysis since we only measure the average mass of a set of clusters.
}

While CMB polarization data, which are almost unaffected by the presence of foregrounds, are expected to provide robust lensing estimates with the future low-noise CMB datasets like CMB-S4 \citepalias{raghunathan17a}, the estimator presented here and in a similar work by \citet{madhavacheril18} will be vital to extract lensing robustly from future low noise CMB temperature datasets.

\section*{}
We thank useful conversations with Mathew Madhavacheril at the Stanford CMB-lensing meeting in 2017 which partly inspired us in building this modified QE.  
Melbourne group acknowledges support from the Australian Research Council's Discovery Projects scheme (DP150103208). 
SR also acknowledges partial support from the Laby Foundation. 
LB's work was supported under the U.S. Department of Energy contract DE-AC02-06CH11357.
We acknowledge the use of \texttt{HEALPix} \citep{gorski05} and \texttt{CAMB} \citep{lewis00} routines. 

This work was performed in the context of the South Pole Telescope scientific program. SPT is supported by the National Science Foundation through grant PLR-1248097.  Partial support is also provided by the NSF Physics Frontier Center grant PHY-1125897 to the Kavli Institute of Cosmological Physics at the University of Chicago, the Kavli Foundation and the Gordon and Betty Moore Foundation grant GBMF 947. This research used resources of the National Energy Research Scientific Computing Center (NERSC), a DOE Office of Science User Facility supported by the Office of Science of the U.S. Department of Energy under Contract No. DE-AC02-05CH11231. 

Funding for the DES Projects has been provided by the U.S. Department of Energy, the U.S. National Science Foundation, the Ministry of Science and Education of Spain, 
the Science and Technology Facilities Council of the United Kingdom, the Higher Education Funding Council for England, the National Center for Supercomputing 
Applications at the University of Illinois at Urbana-Champaign, the Kavli Institute of Cosmological Physics at the University of Chicago, 
the Center for Cosmology and Astro-Particle Physics at the Ohio State University,
the Mitchell Institute for Fundamental Physics and Astronomy at Texas A\&M University, Financiadora de Estudos e Projetos, 
Funda{\c c}{\~a}o Carlos Chagas Filho de Amparo {\`a} Pesquisa do Estado do Rio de Janeiro, Conselho Nacional de Desenvolvimento Cient{\'i}fico e Tecnol{\'o}gico and 
the Minist{\'e}rio da Ci{\^e}ncia, Tecnologia e Inova{\c c}{\~a}o, the Deutsche Forschungsgemeinschaft and the Collaborating Institutions in the Dark Energy Survey. 

The Collaborating Institutions are Argonne National Laboratory, the University of California at Santa Cruz, the University of Cambridge, Centro de Investigaciones Energ{\'e}ticas, 
Medioambientales y Tecnol{\'o}gicas-Madrid, the University of Chicago, University College London, the DES-Brazil Consortium, the University of Edinburgh, 
the Eidgen{\"o}ssische Technische Hochschule (ETH) Z{\"u}rich, 
Fermi National Accelerator Laboratory, the University of Illinois at Urbana-Champaign, the Institut de Ci{\`e}ncies de l'Espai (IEEC/CSIC), 
the Institut de F{\'i}sica d'Altes Energies, Lawrence Berkeley National Laboratory, the Ludwig-Maximilians Universit{\"a}t M{\"u}nchen and the associated Excellence Cluster Universe, 
the University of Michigan, the National Optical Astronomy Observatory, the University of Nottingham, The Ohio State University, the University of Pennsylvania, the University of Portsmouth, 
SLAC National Accelerator Laboratory, Stanford University, the University of Sussex, Texas A\&M University, and the OzDES Membership Consortium.

Based in part on observations at Cerro Tololo Inter-American Observatory, National Optical Astronomy Observatory, which is operated by the Association of 
Universities for Research in Astronomy (AURA) under a cooperative agreement with the National Science Foundation.

The DES data management system is supported by the National Science Foundation under Grant Numbers AST-1138766 and AST-1536171.
The DES participants from Spanish institutions are partially supported by MINECO under grants AYA2015-71825, ESP2015-66861, FPA2015-68048, SEV-2016-0588, SEV-2016-0597, and MDM-2015-0509, 
some of which include ERDF funds from the European Union. IFAE is partially funded by the CERCA program of the Generalitat de Catalunya.
Research leading to these results has received funding from the European Research
Council under the European Union's Seventh Framework Program (FP7/2007-2013) including ERC grant agreements 240672, 291329, and 306478.
We  acknowledge support from the Australian Research Council Centre of Excellence for All-sky Astrophysics (CAASTRO), through project number CE110001020, and the Brazilian Instituto Nacional de Ci\^encia e Tecnologia (INCT) e-Universe (CNPq grant 465376/2014-2).

This manuscript has been authored by Fermi Research Alliance, LLC under Contract No. DE-AC02-07CH11359 with the U.S. Department of Energy, Office of Science, Office of High Energy Physics. The United States Government retains and the publisher, by accepting the article for publication, acknowledges that the United States Government retains a non-exclusive, paid-up, irrevocable, world-wide license to publish or reproduce the published form of this manuscript, or allow others to do so, for United States Government purposes.
This manuscript has been authored by Fermi Research Alliance, LLC under Contract No. DE-AC02-07CH11359 with the U.S. Department of Energy, Office of Science, Office of High Energy Physics. The United States Government retains and the publisher, by accepting the article for publication, acknowledges that the United States Government retains a non-exclusive, paid-up, irrevocable, world-wide license to publish or reproduce the published form of this manuscript, or allow others to do so, for United States Government purposes.


\newcommand\JCAP{JCAP}

\bibliographystyle{aasjournal}
\ifdefined\ApJsubmit
\IfFileExists{./spt.bib}
{\bibliography{spt}}
{\bibliography{../../BIBTEX/spt}}
\else
\IfFileExists{./spt.bib}
{\bibliography{spt}}
{\bibliography{../../BIBTEX/spt}}
\fi

\ifdefined\allauthorsinfront
\else
\allauthors
\fi
\end{document}

%% file: principal_authors.tex
\author[0000-0003-1405-378X]{S.~Raghunathan} \affiliation{School of Physics, University of Melbourne, Parkville, VIC 3010, Australia} \affiliation{Department of Physics and Astronomy, University of California, Los Angeles, CA, USA 90095}

\author{S.~Patil} \affiliation{School of Physics, University of Melbourne, Parkville, VIC 3010, Australia}

\author{E.~Baxter} \affiliation{Department of Physics and Astronomy, University of Pennsylvania, Philadelphia, PA 19104, USA}

\author[0000-0002-5108-6823]{B.~A.~Benson} \affiliation{Fermi National Accelerator Laboratory, MS209, P.O. Box 500, Batavia, IL 60510} \affiliation{Kavli Institute for Cosmological Physics, University of Chicago, 5640 South Ellis Avenue, Chicago, IL, USA 60637} \affiliation{Department of Astronomy and Astrophysics, University of Chicago, 5640 South Ellis Avenue, Chicago, IL, USA 60637}

\author{L.~E.~Bleem} \affiliation{High Energy Physics Division, Argonne National Laboratory, 9700 S. Cass Avenue, Argonne, IL, USA 60439} \affiliation{Kavli Institute for Cosmological Physics, University of Chicago, 5640 South Ellis Avenue, Chicago, IL, USA 60637}

\author[0000-0002-3091-8790]{T.~L.~Chou} \affiliation{Department of Physics, University of Chicago, Chicago, IL, USA 60637}

\author{T.~M.~Crawford} \affiliation{Kavli Institute for Cosmological Physics, University of Chicago, 5640 South Ellis Avenue, Chicago, IL, USA 60637} \affiliation{Department of Astronomy and Astrophysics, University of Chicago, 5640 South Ellis Avenue, Chicago, IL, USA 60637}

\author{G.~P.~Holder} \affiliation{Department of Physics, McGill University, 3600 Rue University, Montreal, Quebec H3A 2T8, Canada} \affiliation{Canadian Institute for Advanced Research, CIFAR Program in Cosmology and Gravity, Toronto, ON, M5G 1Z8, Canada}

\author{T.~McClintock} \affiliation{Department of Physics, University of Arizona, Tucson, AZ 85721, USA}

\author{C.~L.~Reichardt} \affiliation{School of Physics, University of Melbourne, Parkville, VIC 3010, Australia}

\author{E.~Rozo} \affiliation{Department of Physics, University of Arizona, Tucson, AZ 85721, USA}

\author{T.~N.~Varga} \affiliation{Max Planck Institute for Extraterrestrial Physics, Giessenbachstrasse, 85748 Garching, Germany} \affiliation{Universit\"ats-Sternwarte, Fakult\"at f\"ur Physik, LudwigMaximilians Universit\"at M\"unchen, Scheinerstr. 1, 81679 M\"unchen, Germany}

%% file: spt_des_authorlist.tex
\author{T.~M.~C.~Abbott} \affiliation{Cerro Tololo Inter-American Observatory, National Optical Astronomy Observatory, Casilla 603, La Serena, Chile}

\author{P.~A.~R.~Ade} \affiliation{Cardiff University, Cardiff CF10 3XQ, United Kingdom}

\author[0000-0002-7069-7857]{S.~Allam} \affiliation{Fermi National Accelerator Laboratory, P. O. Box 500, Batavia, IL 60510, USA}

\author{A.~J.~Anderson} \affiliation{Fermi National Accelerator Laboratory, MS209, P.O. Box 500, Batavia, IL 60510}

\author[0000-0002-0609-3987]{J.~Annis} \affiliation{Fermi National Accelerator Laboratory, P. O. Box 500, Batavia, IL 60510, USA}

\author{J.~E.~Austermann} \affiliation{NIST Quantum Devices Group, 325 Broadway Mailcode 817.03, Boulder, CO, USA 80305}

\author{S.~Avila} \affiliation{Institute of Cosmology \& Gravitation, University of Portsmouth, Portsmouth, PO1 3FX, UK}

\author{J.~A.~Beall} \affiliation{NIST Quantum Devices Group, 325 Broadway Mailcode 817.03, Boulder, CO, USA 80305}

\author{K.~Bechtol} \affiliation{LSST, 933 North Cherry Avenue, Tucson, AZ 85721, USA}

\author{A.~N.~Bender} \affiliation{High Energy Physics Division, Argonne National Laboratory, 9700 S. Cass Avenue, Argonne, IL, USA 60439} \affiliation{Kavli Institute for Cosmological Physics, University of Chicago, 5640 South Ellis Avenue, Chicago, IL, USA 60637}

% ...
